\documentclass[final,1p,times]{elsarticle} % do not change
\usepackage{graphicx} % do not change
\usepackage{amssymb} % do not change
\usepackage{amsthm} % do not change
\usepackage{lineno} % do not change

\journal{Nuclear Physics A} % do not change
\begin{document} % do not change

\begin{frontmatter} % do not change

%% QM09Author: please enter your  
%% Title, author and address info here; please do not use footnotes

% Your Title - please insert
\title{The centrality dependence of $v_2/\varepsilon$: the ideal hydro limit and $\eta/s$}

% Principle author, and co-authors - please insert
\author{H.~Masui, J-Y.~Ollitrault, \underline{R.~Snellings}, A.~Tang}

% Address - please insert
\address{}

\begin{abstract} % do not change
%% Text of abstract goes here - please insert
The large elliptic flow observed at {\sc RHIC} is considered to be evidence for almost perfect liquid behavior of the strongly coupled quark-gluon plasma produced in the collisions. 
In these proceedings we present a two parameter fit for the centrality dependence of the elliptic flow $v_2$ scaled by the spatial eccentricity $\varepsilon$. 
We show by comparing to viscous hydrodynamical calculations that these two parameters are in good approximation proportional to the shear viscosity over entropy ratio $\eta/s$ and the ideal hydro limit of the ratio $v_2/\varepsilon$.
\end{abstract} % do not change

\end{frontmatter} % do not change

%% QM09: we keep linenumbers at least for initial version
%\linenumbers % do not change

%% start of main text - please insert. 
\section{Introduction}
\label{Introduction}
The goal of the ultra-relativistic nuclear collision program is the
creation and study of a new state of matter, the quark-gluon plasma. The
azimuthal anisotropy of the transverse momentum distribution in 
non-central heavy-ion collisions is thought to be sensitive to the
properties of this state of matter. The second Fourier coefficient of this
anisotropy, $v_{2}$, is called elliptic flow. 
For a recent review see~\cite{Voloshin:2008dg}.

In ideal hydrodynamics $v_2$ is proportional to the spatial eccentricity 
with a magnitude which depends on the Equation of State {\sc EoS}.
This spatial eccentricity is defined by
\[
\varepsilon = \frac{\langle y^2 - x^2 \rangle}{\langle y^2 + x^2 \rangle}
\]
where $x$ and $y$ are the spatial coordinates of the colliding nucleons in the plane
perpendicular to the collision axis and where the brackets 
denote an average.
In practice $\varepsilon$ is not a measured quantity but obtained from model calculations, 
using Glauber or Color Glass Condensate ({\sc CGC}) models, for instance.

The ratio $v_2/\varepsilon$ versus particle density is a sensitive gauge to test if the system approaches  ideal hydrodynamic behavior~\cite{Bhalerao:2005mm}. It was observed that this ratio reaches the expected ideal 
hydrodynamic values only for the more central collisions at the highest RHIC center of mass 
energy~\cite{Alt:2003ab,Adler:2002pu} which indicates that certainly for non-central collisions, as well as at lower energies, and away from mid-rapidity the elliptic flow contains significant non-ideal hydro contributions. 

Much of this discrepancy can be explained by incorporating the viscous contribution from the 
hadronic phase~\cite{Teaney:2001av,Hirano:2005wx,Hirano:2005xf}.
However, we expect that also the hot and dense phase must deviate from an 
ideal hydrodynamic description. 
Kovtun, Son and Starinets ({\sc KSS})~\cite{Kovtun:2004de}, showed that conformal field theories 
with gravity duals have a ratio of shear viscosity $\eta$ to entropy density $s$ of,  in natural units, $\eta/s = 1/4\pi$.
They conjectured that this value is a lower bound for any relativistic thermal field theory. 
In addition, Teaney~\cite{Teaney:2003kp} pointed out that very small shear viscosities, 
of the magnitude of the bound, would already lead to a significant reduction in the predicted elliptic flow. 

Based on the centrality dependence of $v_2/\varepsilon$, the magnitude of $\eta/s$ for the created system has been estimated recently from a transport theory motivated calculation~\cite{Drescher:2007cd,Liu:2009zz} and 
from viscous hydrodynamical calculations~\cite{Song:2008si, Luzum:2008cw}.
Both approaches have their merits and drawbacks. 

In these proceedings we explore how well a parameterization can be used to estimate $\eta/s$ as well as the ideal hydrodynamical limit of $v_2/\varepsilon$  which is closely related to the EoS.

\begin{figure}[b!]
  \begin{minipage}[t]{0.49\textwidth}
    \includegraphics[width=1.\textwidth]{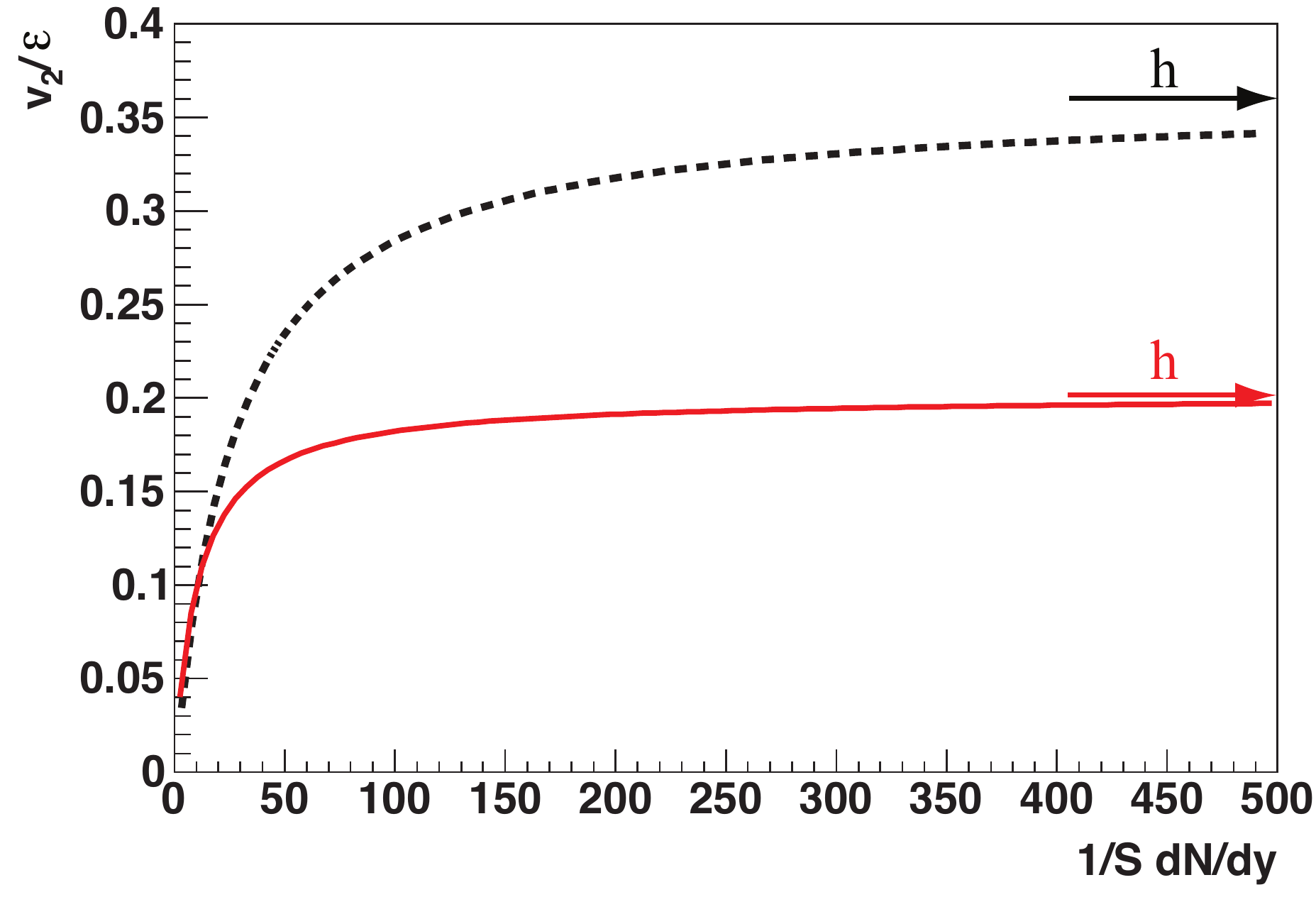}
    \caption{The dependence of $v_2/\varepsilon$ versus transverse density 
    of equation~\ref{parameterization1} for two values of $h$ and two values of $\eta/s$.}
    \label{fitexample}
  \end{minipage}
  \hspace{\fill}
  \begin{minipage}[t]{0.49\textwidth}
    \includegraphics[width=1.\textwidth]{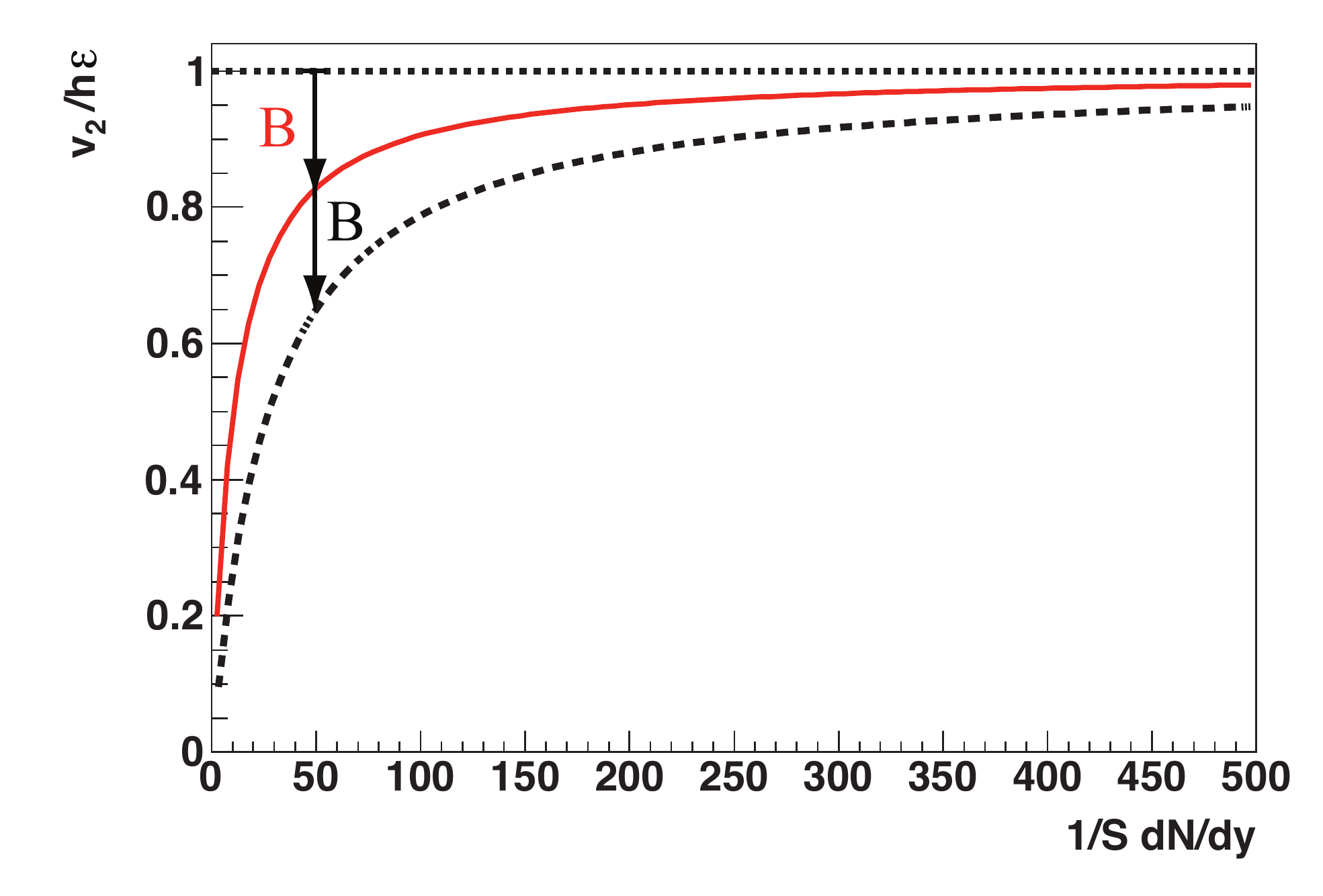}
    \caption{The dependence of $v_2/ h\varepsilon$ versus transverse density of equation~\ref{parameterization1} for the same parameters as Fig.~\ref{fitexample}.}
    \label{fitexamplenorm}
\end{minipage}
\end{figure}
\section{Simple Parameterization}
\label{Fitting}
We use the parameterization from~\cite{Bhalerao:2005mm,Drescher:2007cd} which is defined  by
\begin{equation}
\frac{v_2}{\varepsilon} = \frac{h}{1+B/(\frac{1}{S}\frac{dN}{dy})},
\label{parameterization1}
\end{equation}
where $S$ is the transverse area of the collision region and $h$ and $B$ are the two free parameters
in the fit.
The parameter $h$ corresponds to the ideal hydro limit of $v_2/\varepsilon$ and $B$ is proportional 
to $\eta/s$.

Figure~\ref{fitexample} shows how the parameterization behaves for two different values of the ideal
 hydro limit (the dashed line represents the harder EoS) and two different values 
 of $\eta/s$ (the full line represents the smaller $\eta/s$). 
 The effect of the EoS is clearly seen in the magnitude of $v_2/\varepsilon$ in Fig.~\ref{fitexample} and 
 the value of $\eta/s$ is reflected by the change in this magnitude versus $1/S dN/dy$ (for $\eta/s = 0$ the
 magnitude will be constant).  
 The magnitude of $\eta/s$ is easier to quantify if one plots $v_2/h\varepsilon$, as is done in Fig.~\ref{fitexamplenorm}. A larger deviation from unity at fixed value of $1/S dN/dy$ then indicates a larger $\eta/s$.

\begin{figure}[t!]
  \begin{minipage}[t]{0.49\textwidth}
    \includegraphics[width=1.\textwidth]{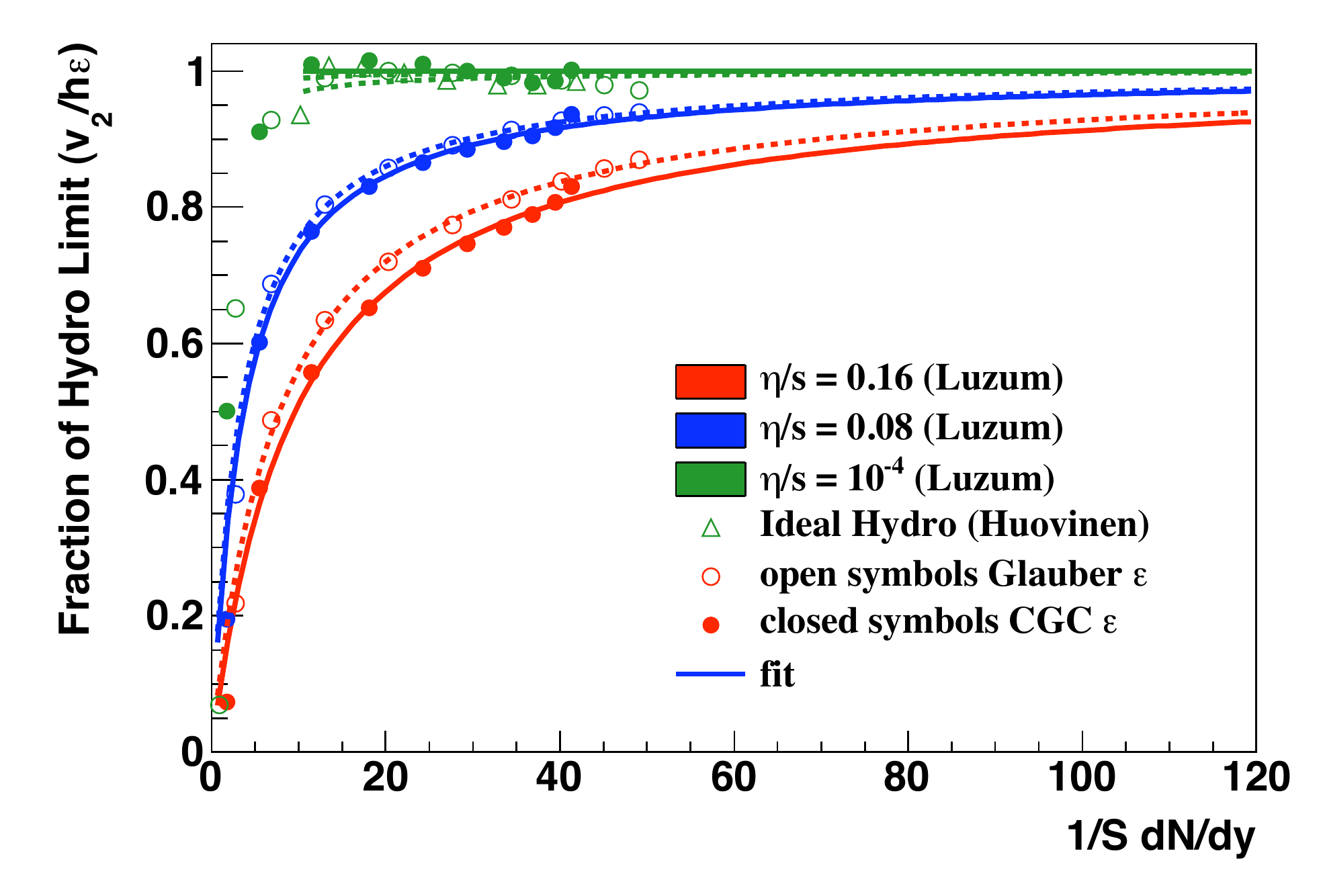}
    \caption{A fit of viscous hydrodynamical model results using CGC and Glauber initial eccentricities 
    with Eq.~\ref{parameterization1}.}
    \label{compare_epsilon}
  \end{minipage}
  \hspace{\fill}
  \begin{minipage}[t]{0.49\textwidth}
    \includegraphics[width=1.\textwidth]{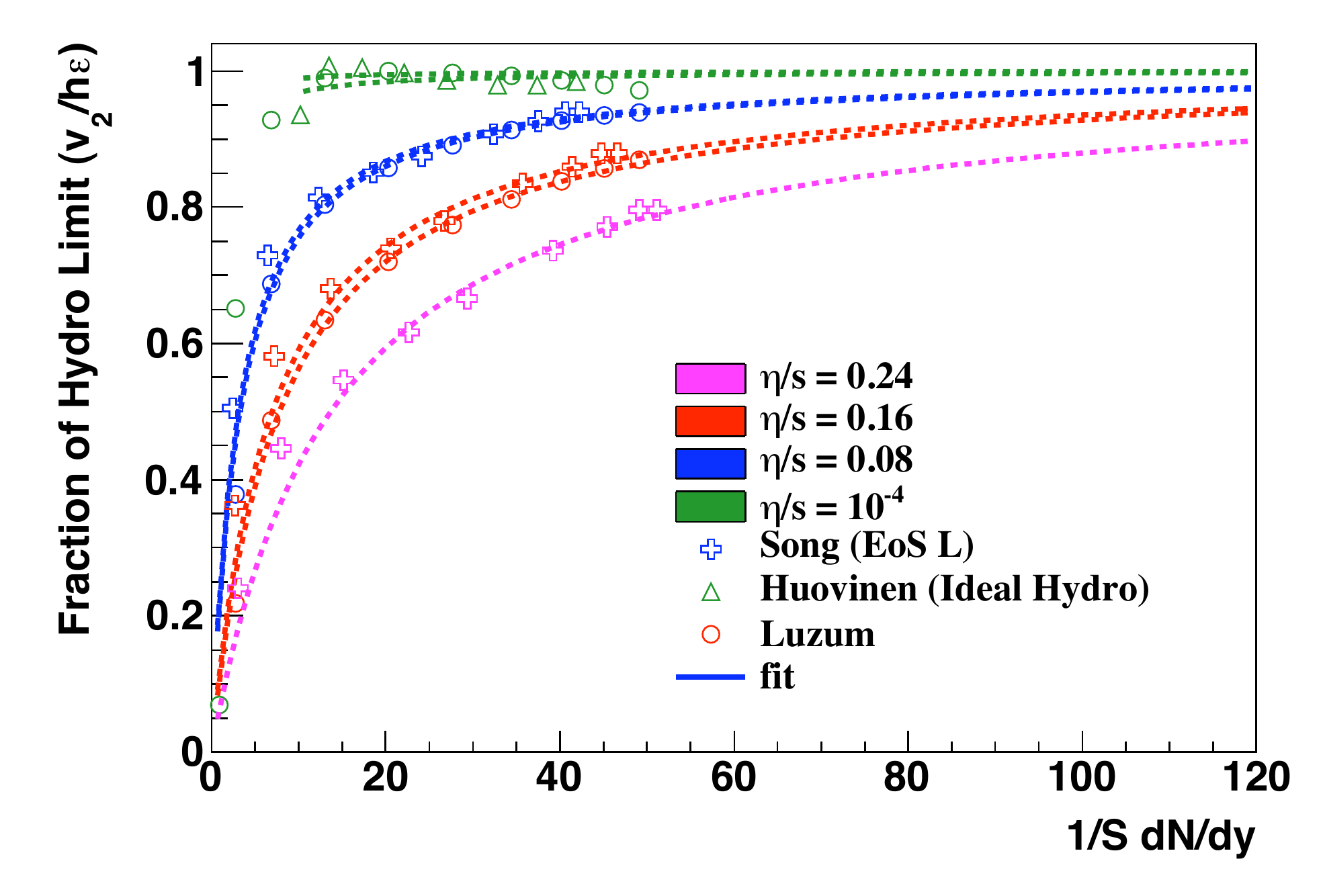}
    \caption{Comparing viscous hydrodynamical calculations of different groups with the fit}
    \label{compare_theorist}
\end{minipage}
\end{figure}
To test if this simple parameterization does describe a state of the art viscous hydrodynamical calculation 
we fit the calculations from Luzum and Romatschke~\cite{Luzum:2009sb}. 
Figure~\ref{compare_epsilon} shows that Eq.~\ref{parameterization1} well describes results from viscous hydrodynamical calculations, done with three different values of $\eta/s$ and two different parameterizations of the spatial eccentricity (Glauber and CGC). 
As expected, $v_2$ is to good approximation proportional to 
the initial spatial eccentricity.  
In addition, it is seen that the deviation of $v_2/\varepsilon$ from unity at a given $1/S dN/dy$ increases for larger values of $\eta/s$. 

Figure~\ref{compare_theorist} shows $v_2/h\varepsilon$ from viscous hydrodynamical calculations~\cite{Song:2008si,Luzum:2008cw,Luzum:2009sb} 
done by different groups using the same set of values of $\eta/s$ but different parameterization 
of the {\sc EoS} and $\varepsilon$. The value of $\varepsilon$ is that used in the hydrodynamical
calculations while the value of $h$ is obtained from the fit.
We conclude that our parameterization yields curves that depend on the value of $\eta/s$ but are roughly 
independent of the {\sc EoS} and $\varepsilon$.
However it turns out that if the EoS is very different (e.g. not incorporating a phase transition) this scaling 
does break down (not shown).

\begin{figure}[b!]
  \begin{minipage}[t]{0.49\textwidth}
    \includegraphics[width=1.\textwidth]{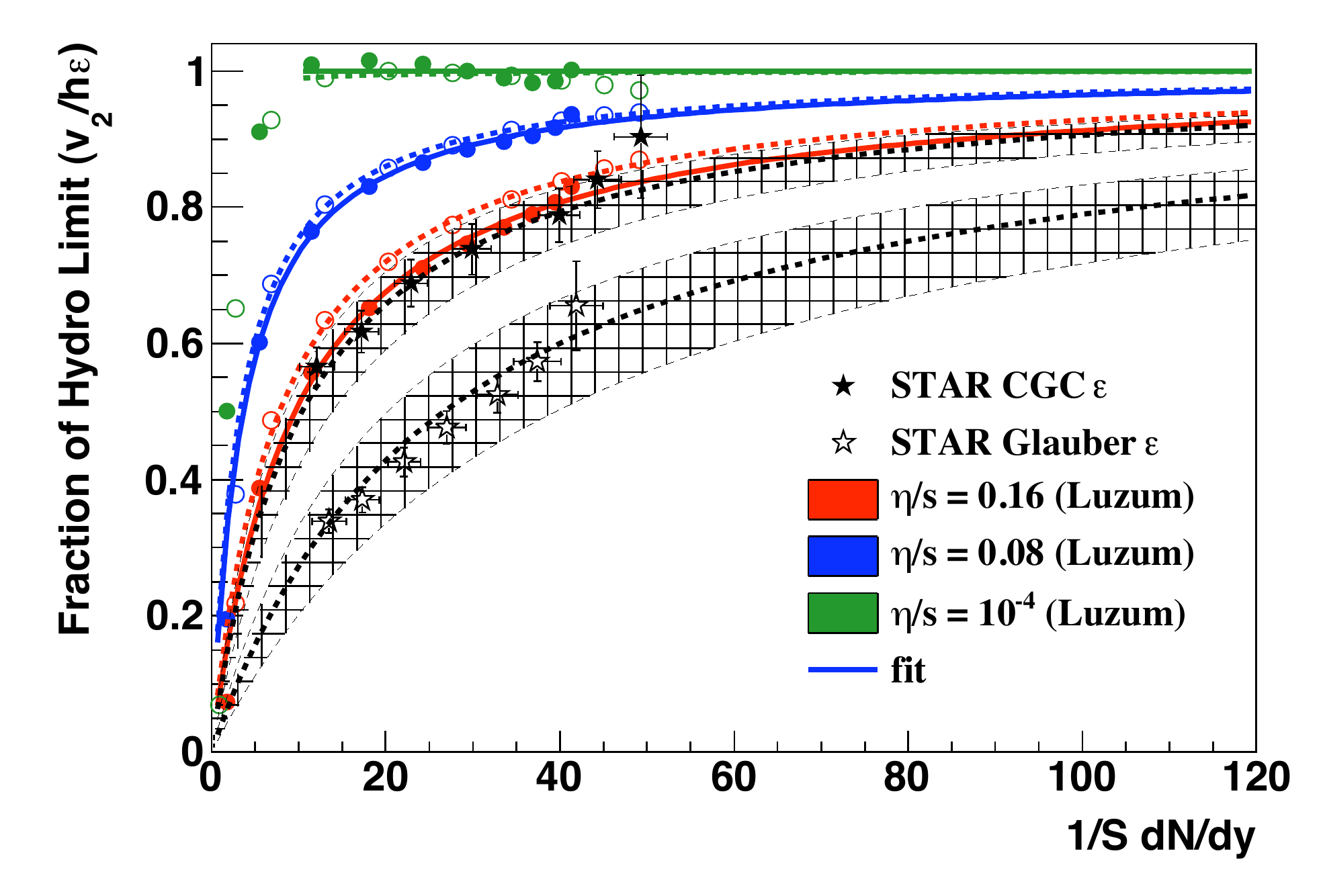}
    \caption{Comparing viscous hydrodynamical calculations with STAR data.}
    \label{compare_data}
  \end{minipage}
  \hspace{\fill}
  \begin{minipage}[t]{0.49\textwidth}
    \includegraphics[width=1.\textwidth]{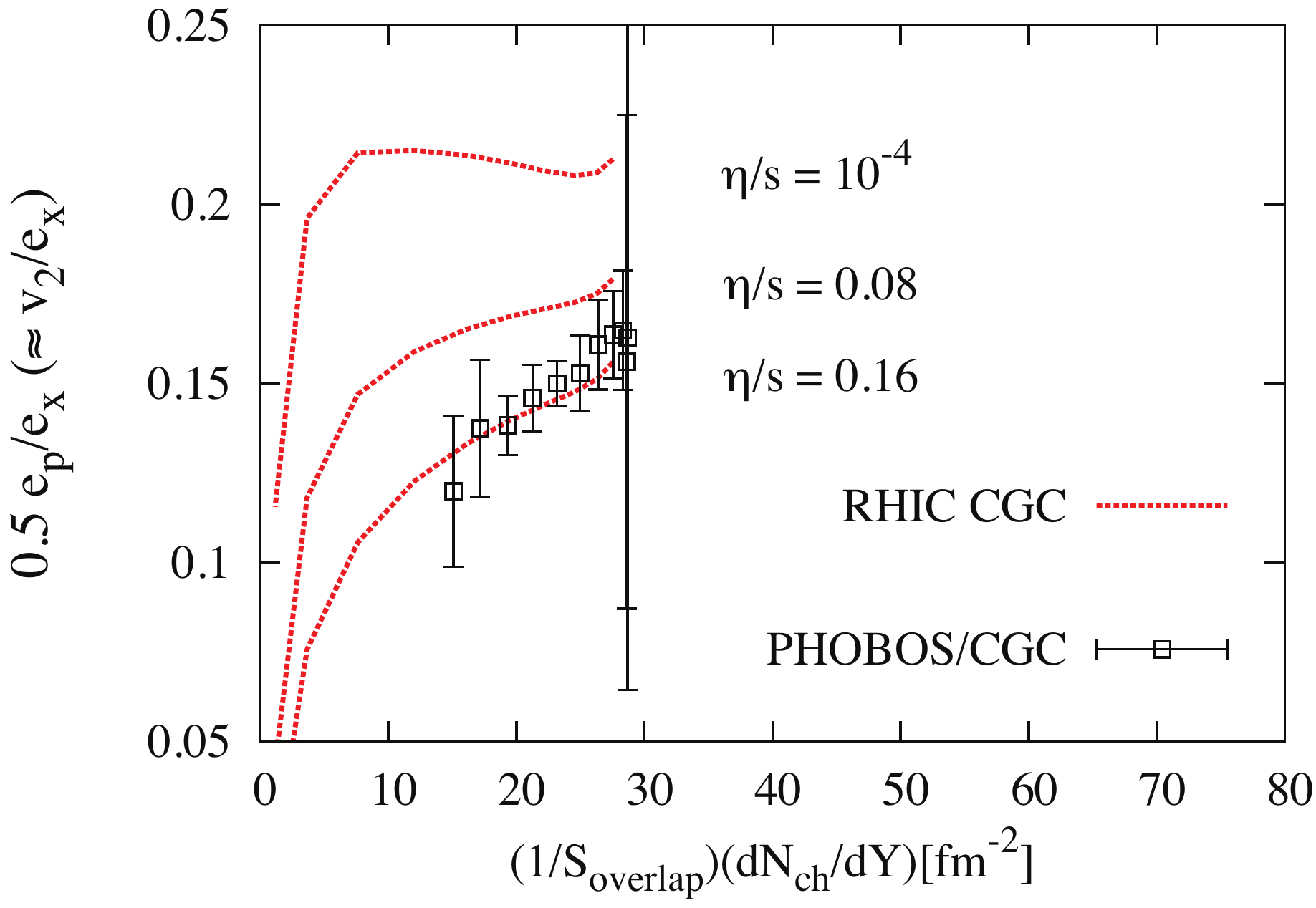}
    \caption{A direct comparison of viscous hydro calculations with PHOBOS data 
    (from~\cite{Luzum:2009sb}).}
    \label{v2viscous_hydroCGC}
\end{minipage}
\end{figure}
Using Eq.~\ref{parameterization1}, we can now compare the various viscous hydrodynamical results with data 
and estimate the value of $\eta/s$. 
Since the value of $\varepsilon$ is not known we take the
eccentricity calculated assuming 
CGC~\cite{Drescher:2006pi} or Glauber (wounded nucleon) initial conditions
as two extremes.
It is seen from Fig.~\ref{compare_data} that, assuming the CGC initial conditions, 
the STAR data is well described with twice the KSS bound, $\eta/s = 2/4\pi$. 
Using the Glauber initial conditions, however, the STAR data is not described 
within the range of $\eta/s$ currently used by the viscous hydrodynamic calculations. 
From the deviation from unity one can estimate that the corresponding value of $\eta/s$ would be 
approximately four times the KSS bound.
Using the CGC or Glauber initial conditions we find for the ideal hydro limit of $v_2/\varepsilon$ the value 
$0.20 \pm 0.0$1 and $0.36 \pm 0.07$, respectively. 

For the CGC initial conditions the value of $h$ approximately matches the EoS used by 
Luzum and Romatschke~\cite{Luzum:2008cw}). 
This is illustrated in Fig.~\ref{v2viscous_hydroCGC} where the centrality 
dependence of $v_2$~\cite{Luzum:2009sb} is well described by {\sc CGC} initial conditions,
a value of $h \approx 0.2$ and $\eta/s = 2/4\pi$.

\begin{figure}[t!]
  \begin{minipage}[t]{0.63\textwidth}
    \includegraphics[width=1.\textwidth]{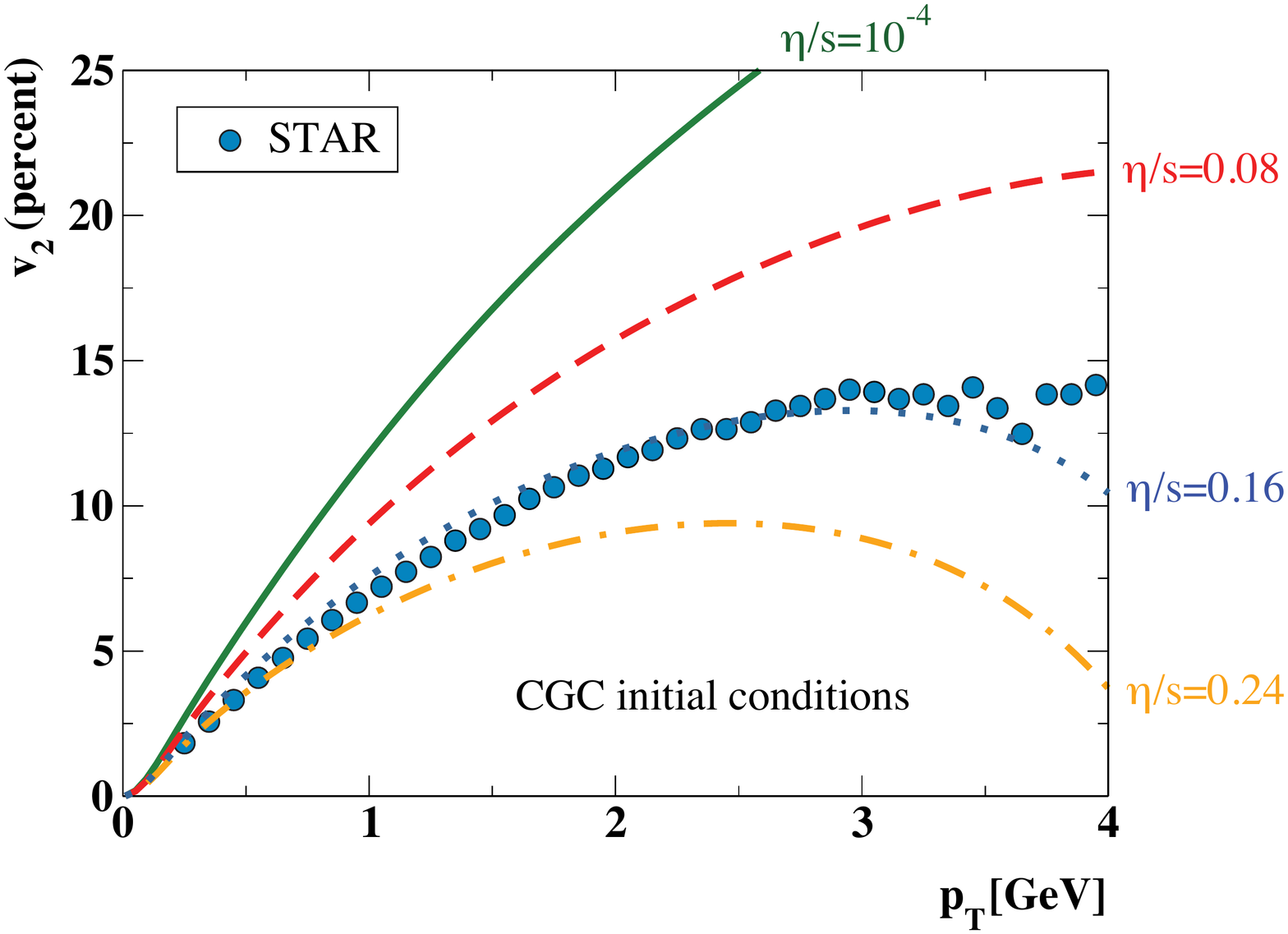}
    \caption{$v_2$ from STAR (approximately corrected for nonflow) compared to 
    viscous hydrodynamical calculations (from~\cite{Luzum:2008cw}).}
    \label{v2ptviscous_hydroCGC}
  \end{minipage}
  \hspace{\fill}
  \begin{minipage}[t]{0.34\textwidth}
    \includegraphics[width=1.\textwidth]{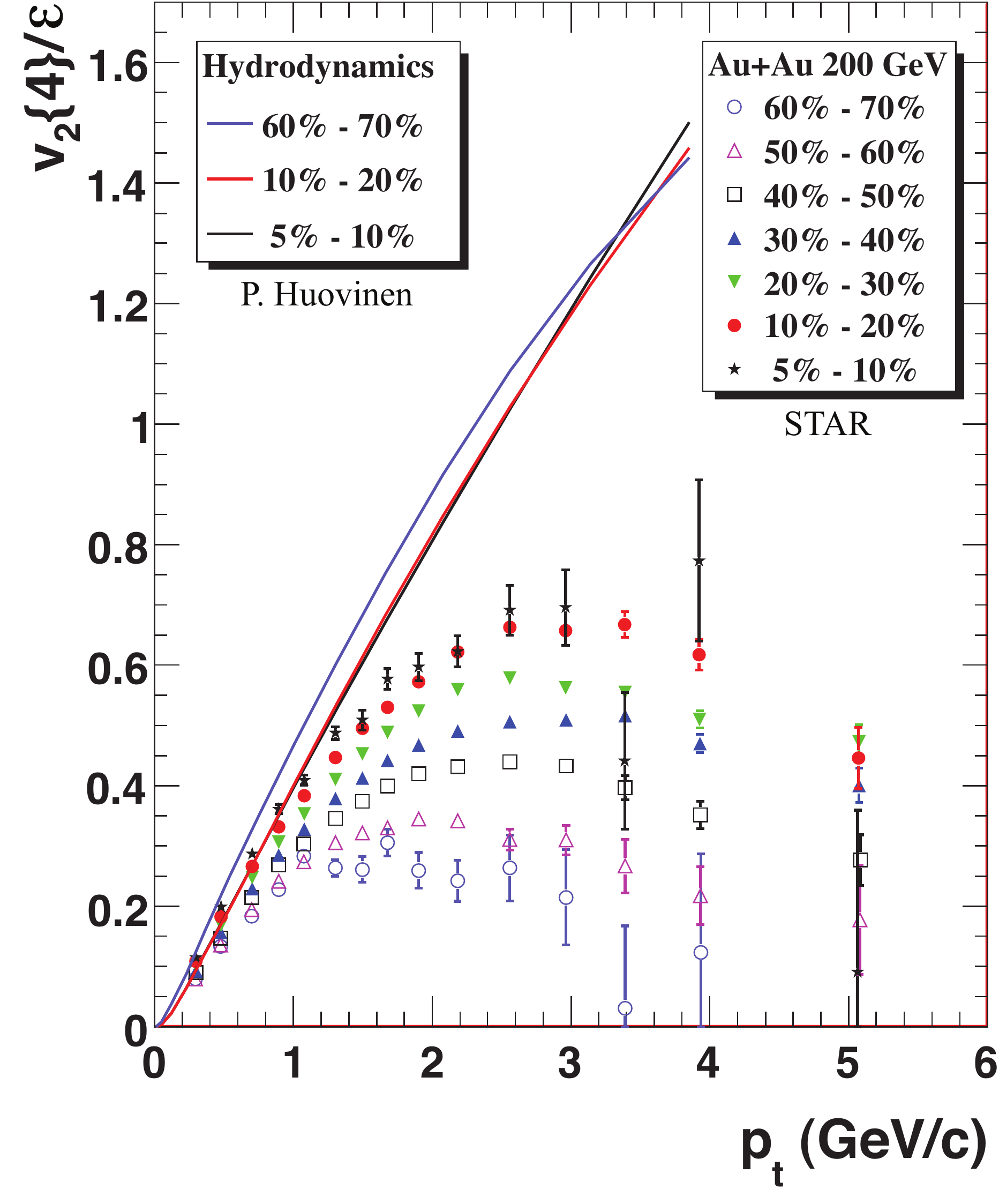}
    \caption{$v_2$ from STAR as function of transverse momentum and centrality.}
    \label{v2pt_STAR}
\end{minipage}
\end{figure}

Using viscous hydrodynamics with these CGC initial conditions, {\sc EoS}, and magnitude of $\eta/s$, 
the transverse momentum dependence of $v_2$ is also well described, 
as shown in Fig.~\ref{v2ptviscous_hydroCGC}. 
The figure illustrates that the $p_{\mathrm t}$ dependence is very sensitive to the viscous correction such that larger
corrections decrease the magnitude of $v_2$ and shift its maximum to lower $p_{\mathrm t}$.
Figure~\ref{v2pt_STAR} shows the centrality dependence of $v_2(p_{\mathrm t})$ where one clearly observes that 
the deviation with $\eta/s = 0$ increases from central to peripheral collisions and that the peak position shifts
to lower $p_{\mathrm t}$, consistent with larger viscous effects.
  
\section{Conclusions}
\label{Conclusions}
We have shown that a simple parameterization can describe the centrality dependence of $v_2/\varepsilon$. 
When compared to viscous hydrodynamical calculations such a parameterization yields an estimate of $\eta/s$.
We find that the current RHIC data is described well by a spatial eccentricity based on CGC initial conditions,
a soft EoS with $v_2/\varepsilon \approx 0.2$ and $\eta/s$ twice the KSS bound.

%% end of main text

%\section*{Acknowledgments} % please insert, comment out or delete if not needed
%This is where one places acknowledgments for funding bodies etc., if needed.
%For the large collaborations, this is listed once and for all, together with 
%the author lists etc. in the proceedings back-material.

 % do not change 
\end{document}